\documentclass[10pt,twocolumn,letterpaper]{article}

\usepackage[utf8]{inputenc}
\usepackage[T1]{fontenc}
\usepackage{mathptmx}                 
\usepackage[scaled=0.92]{helvet}
\usepackage{courier}
\usepackage{microtype}
\usepackage[letterpaper,top=0.75in,bottom=1in,left=0.75in,right=0.75in,columnsep=0.25in]{geometry}
\usepackage{amsmath,amssymb}
\usepackage{graphicx}
\usepackage{booktabs}
\usepackage{tabularx}
\usepackage{array}
\usepackage{ragged2e}
\usepackage{rotating}
\usepackage{caption}
\usepackage{enumitem}
\usepackage{titlesec}
\usepackage{authblk}
\usepackage{xcolor}
\usepackage[style=ieee, doi=true]{biblatex}   
\PassOptionsToPackage{hyphens}{url}
\usepackage[hidelinks,breaklinks]{hyperref}
\usepackage{url}

\renewcommand{\thesection}{\Roman{section}}
\renewcommand{\thesubsection}{\Alph{subsection}}
\renewcommand{\thesubsubsection}{\arabic{subsubsection}}
\titleformat{\section}{\centering\normalsize\scshape}{\thesection.}{0.5em}{}
\titleformat{\subsection}{\normalsize\itshape}{\thesubsection.}{0.5em}{}
\titleformat{\subsubsection}[runin]{\normalsize\itshape}{\thesubsubsection)}{0.5em}{}[:]
\titlespacing*{\section}{0pt}{1.4em}{0.8em}
\titlespacing*{\subsection}{0pt}{1.0em}{0.5em}
\titlespacing*{\subsubsection}{0pt}{0.7em}{0.5em}

\newcolumntype{L}{>{\RaggedRight\arraybackslash}X}

\setlist{nosep,leftmargin=*}

\title{\vspace{-1.5em}{\normalsize\textsc{Study Protocol}}\\[0.6em]
\LARGE Simulating access to alcohol screening, brief intervention, referral to and receipt of specialty treatment for alcohol use disorder in California and Texas: A study protocol for an agent-based modeling study (AUDTxSim)}

\author[1]{Rainer Hilscher}
\author[1]{Andy Kawabata}
\author[1]{Nick Kruskamp}
\author[1]{William Dowd}
\author[2]{Nina Mulia}
\author[1]{Carolina Barbosa}
\author[1]{Georgiy Bobashev}
\author[1]{Jessica Duncan Cance}
\author[1]{Ivette Rodriguez Borja}
\author[1]{James I. Rineer}
\author[1]{Katherine J. Karriker-Jaffe}
\affil[1]{RTI International, Research Triangle Park, NC, USA}
\affil[2]{Alcohol Research Group, Public Health Institute, Berkeley, CA, USA}
\date{}

\begin{document}
\maketitle

\begingroup\small\noindent\textbf{\textit{Abstract}---}Alcohol use disorder (AUD) affects millions of adults in the United States, contributing to substantial burden on individuals, their families, and the health care systems that support them. Primary care--based screening, brief intervention, and referral to treatment (SBIRT) is a leading strategy for identifying heavy alcohol use before it progresses, but access to primary care, screening, and downstream specialty alcohol treatment differs markedly across population subgroups. Whether scaling SBIRT and expanding treatment capacity narrows or widens these differences at the population level is not known and cannot feasibly be answered with a randomized trial. In this protocol we describe AUDTxSim (AUD Treatment Simulation), a stochastic, spatially explicit, individual-based simulation model of alcohol use, AUD, SBIRT, and specialty treatment for the adult populations of California and Texas. Agents represent every adult in synthetic populations of the two states (approximately 28.0 million in California and 19.3 million in Texas), each geo-located to a Census tract and linked to the alcohol treatment facilities within reasonable proximity. The model advances in daily time steps and simulates: (1) annual natural-history transitions among eight alcohol use and AUD states; (2) healthcare visits, alcohol screening, brief intervention, and referral to treatment; (3) treatment seeking, standalone medication for AUD (MAUD), mutual help, and requests for hospitalization, residential, or outpatient treatment; (4) a system-level treatment manager that allocates individuals to facilities according to preference, geography, and capacity, with telehealth substitution and waitlists when capacity is unavailable; and (5) treatment progression, stepdown between levels of care, and post-treatment mutual help. The model is implemented in Python with Repast4Py and runs in parallel on a high-performance computing cluster. Once calibrated to state, rural--urban, and demographic-stratum targets for healthcare visits, heavy drinking and AUD prevalence, and mortality, AUDTxSim will be used to project the population-level and subgroup-specific effects of alternative SBIRT reach and treatment capacity scenarios on AUD prevalence, treatment receipt, and mortality. We describe the model structure, decision logic, data sources, technical implementation, and calibration plan so readers can understand what the model will do once calibrated and how its outputs should be interpreted.\par\endgroup
\bigskip

\section{Introduction}

\subsection{The burden of alcohol use disorder and the treatment gap}

According to the 2023 National Survey on Drug Use and Health (NSDUH), 28.1 million United States (US) adults had an alcohol use disorder (AUD) \cite{SAMHSA2022}, but less than 8\% received treatment in the past year. Excessive alcohol use is linked to 25 chronic diseases \cite{Chartier2014} and more than 50 causes of death \cite{Chartier2013}. Heavy alcohol use significantly increases the risk of AUD, liver cirrhosis and liver failure, and multiple cancers, including of the mouth, throat, and liver \cite{Delker2016,Guerrero2013,Iwamoto2016}. In 2023, 12,429 people were killed in motor vehicle crashes involving alcohol-impaired drivers, and alcohol is involved in roughly one-third of all traffic-related deaths in the US \cite{Spillane2007}. The toll of alcohol-related harms is substantial, and this high burden underscores the importance of health care systems as a pivotal entry point for early detection and meaningful intervention.

\subsection{Primary care--based SBIRT and unequal access}

Primary care--based alcohol screening is a public health intervention strategy for reducing heavy alcohol drinking, AUD, and related morbidity and mortality in the general population. The primary care setting offers frequent, low-barrier opportunities to identify risky drinking behaviors before they escalate into more serious health outcomes. Integrating alcohol screening into routine health care also allows providers to deliver a brief intervention that informs patients about the harms and benefits of reducing alcohol intake and offers advice on how to reduce consumption. There is evidence that screening, brief intervention, and referral to treatment (SBIRT) is associated with reduced alcohol consumption, yet demographic differences in access to and receipt of alcohol screening ultimately limit improvements in overall population health \cite{Ballesteros2004,Chi2022,McQueen2011}.

Differences in access to primary care among US demographic subgroups are well documented. Black, Hispanic, and uninsured individuals who are at-risk drinkers (i.e., who exceed recommended drinking guidelines \cite{NIAAA2005}) or who have AUD experience less primary care access than their White or insured counterparts \cite{Mulia2011}. These differences may be explained in part by structural issues such as primary care deserts \cite{Brinzac2023} (residential areas with few or no primary care providers), transportation challenges, and other factors that limit health care availability for certain populations \cite{Buchmueller2005,Caraballo2022,Derose2009}. Foreign-born immigrants also face barriers to primary care because they often lack insurance coverage, and limited English proficiency is associated with lower access and worse quality of care \cite{Lu2020,Pippins2007,Ponce2006}. Consequently, relying on SBIRT in primary care to provide alcohol counseling and to identify individuals in need of specialty alcohol treatment may have the unintended consequence of widening gaps in care receipt and in long-term alcohol severity between demographic groups \cite{Mulia2011}.

\subsection{Simulation modeling can quantify the population impact of expanding SBIRT and treatment}

A randomized trial that allocates entire state populations to different levels of SBIRT reach or treatment capacity is not feasible. Complex systems science offers computational tools that emulate real-world heterogeneity so that such counterfactuals can be examined \emph{in silico}. Agent-based and individual-based models create computer representations of a system comprising heterogeneous individuals, their attributes and histories, and the environment---here, the health care system and the network of specialty treatment facilities. Each simulated individual is endowed with characteristics and probabilities grounded in empirical data, transitions between alcohol use and AUD states over time, and interacts with an environment whose capacity is finite and spatially distributed. Because these models are stochastic, repeated runs reveal the range of plausible population outcomes under a set of specified conditions, and because they carry the full demographic and geographic detail of the population, they permit assessment of outcomes for population subgroups defined by race and ethnicity, sex, age, insurance, language, and rurality \cite{Apostolopoulos2017,Atkinson2017,Wallace2015}.

Models are most useful when tailored to a specific population and question. Our team has previously developed synthetic-population-based microsimulation and agent-based models of opioid overdose, HIV, COVID-19, and healthcare network dynamics \cite{Bates2018,Bobashev2017,Bobashev2020,Jones2019,Rhea2020}. The present protocol describes the design and development of AUDTxSim (AUD Treatment Simulation), a stochastic, spatially explicit, individual-based model of alcohol use, AUD, SBIRT, and specialty treatment for the adult populations of California (CA) and Texas (TX). The goal of this manuscript is to describe the model in sufficient detail that readers understand what the model will do once calibrated, what questions it can answer, and how its outputs should be interpreted.

\section{Materials and Methods}

\subsection{Setting}

AUDTxSim simulates the adult (18 years and older) populations of two large and demographically diverse US states, California and Texas. Together the two states account for roughly one in five US adults and contain large Hispanic, Black, Asian, and foreign-born populations, substantial uninsured and Medicaid-covered populations, and extensive rural areas alongside major metropolitan regions. These features allow the model to represent contrasts in primary care access, screening, and treatment availability between subgroups and between rural and urban settings that are central to the project's overarching goals. Because the two states differ in Medicaid expansion status, insurance coverage, and the density and configuration of specialty treatment facilities, they also provide a natural comparison of how the same intervention strategy may perform under different health-system conditions.

\subsection{Overview of study objectives and simulated scenarios}

The primary purpose of AUDTxSim is to assess the level of access to SBIRT services and the receipt of subsequent specialty treatment---medications for AUD (MAUD), mutual help (e.g., Alcoholics Anonymous and other 12-step programs), and facility-based hospitalization, residential/inpatient, outpatient, and telehealth alcohol treatment---in representative synthetic populations of adults, and to project how alternative levels of SBIRT reach and specialty treatment capacity would change AUD prevalence, treatment receipt, and mortality overall and for key population subgroups. The unit of analysis is the individual. Once the model is calibrated (see Section~\ref{sec:calib}), the following sets of scenarios are planned:

\begin{itemize}
\item \textbf{Scenario A: Status quo (reference).} Current estimated rates of healthcare engagement, alcohol screening, brief intervention, referral, and treatment seeking, with current facility capacity and geography.
\item \textbf{Scenario B: Expanded SBIRT reach.} Increases in the probability of alcohol screening and brief intervention among individuals with healthcare visits (e.g., universal screening), holding treatment capacity constant, to quantify how much of the potential benefit is limited by primary care access and by downstream treatment capacity.
\item \textbf{Scenario C: Expanded treatment capacity and modality.} Increases in facility capacity (via capacity multipliers), expanded telehealth availability, and expanded MAUD availability in both primary care and specialty treatment, with and without expanded SBIRT reach.
\item \textbf{Scenario D: Targeted access improvements.} Scenarios that preferentially increase healthcare engagement or screening among groups with lower baseline access (e.g., uninsured, Hispanic, limited-English-proficiency, or rural residents) to assess which strategies reduce rather than widen subgroup differences.
\end{itemize}

For each scenario the model will produce annual estimates over a five-to-fifteen-year horizon of: prevalence of heavy alcohol use and of mild, moderate, and severe AUD; the proportion of the population screened, receiving a brief intervention, and referred to specialty treatment; the proportion of individuals with AUD who seek treatment, receive standalone MAUD, enter mutual help, and receive facility-based treatment by modality; facility occupancy, waitlist volume, and telehealth substitution; remission and relapse; and all-cause mortality. Each outcome will be stratified by state (California or Texas), rurality, sex, race and ethnicity, age group, insurance status, and county.

\subsection{Model design and conceptual overview}

\begin{figure*}[t]
\centering
\includegraphics[width=\textwidth]{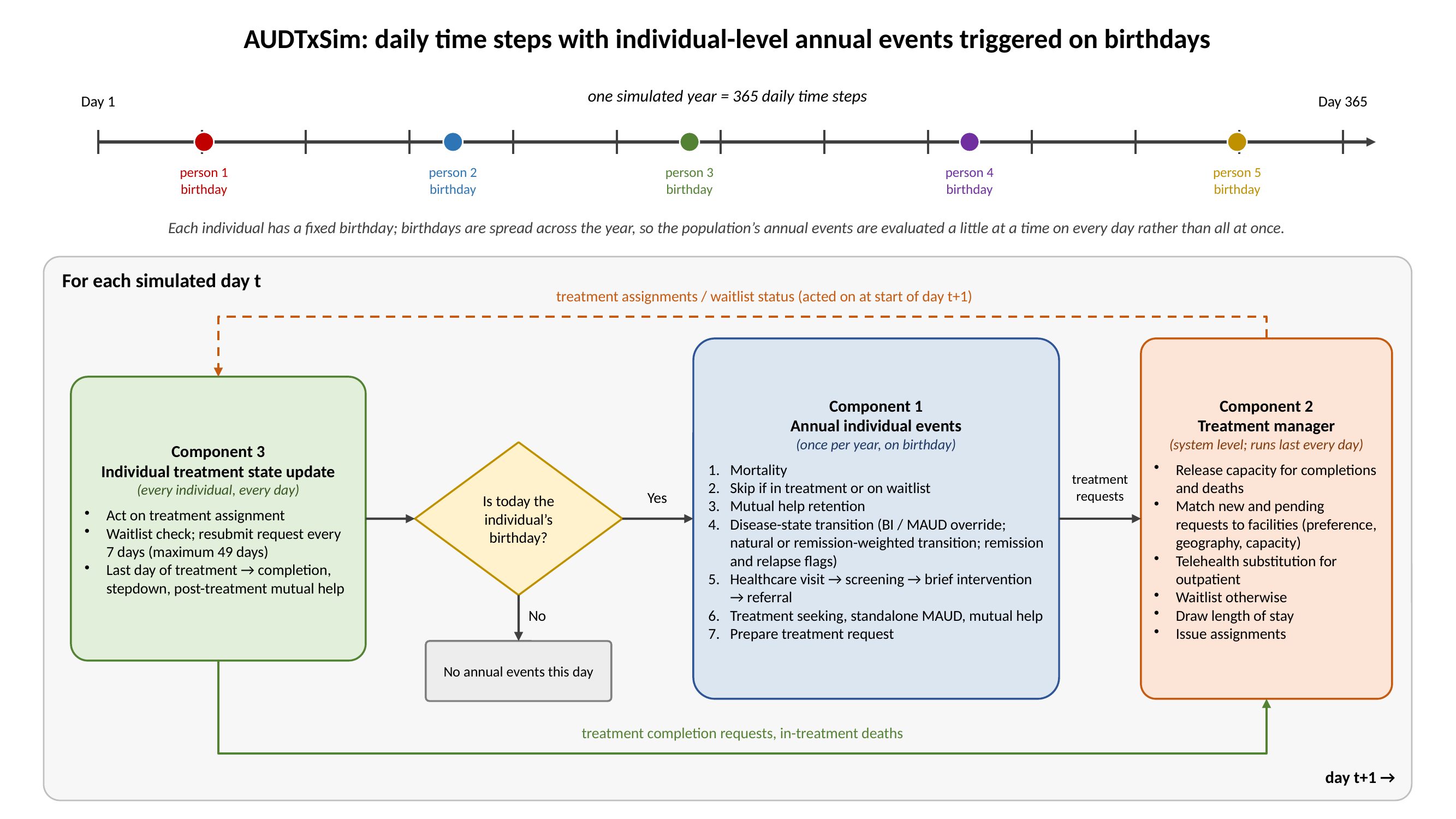}
\caption{Overview of AUDTxSim components, simulated scenarios, and study outcomes.}
\label{fig:overview}
\end{figure*}

AUDTxSim is a stochastic, individual-based model that simulates transitions among alcohol use and AUD states and the interventions that act on them. The model proceeds in daily time steps. Slowly changing individual processes---aging, mortality, natural-history transitions between alcohol use and AUD states, healthcare visits, and SBIRT---are evaluated once per simulated year for each individual on that individual's randomly-assigned birthday, so that the population's annual events are spread evenly across the simulated year. Fast processes---treatment requests, facility allocation, treatment progression, and waitlist handling---are evaluated daily or weekly (waitlist-related repeat treatment requests). Fig.~\ref{fig:overview} summarizes the overall structure.

The model's logic is organized into three interconnected components, which are described in detail in the sections that follow and are drawn in full in Appendix~A (Figs.~\ref{fig:A1}--\ref{fig:A3}):

\begin{itemize}
\item \textbf{Annual individual events and brief intervention} (Fig.~\ref{fig:A1}): birthday, mortality, mutual help retention, natural disease-state transitions, remission and relapse, healthcare visits, screening, brief intervention, referral, treatment seeking, standalone MAUD, and the generation of treatment requests.
\item \textbf{Treatment manager handling of new requests and completions} (Fig.~\ref{fig:A2}): the system-level allocation engine that matches treatment requests to facilities based on treatment type, individual characteristics and preferences, geography, and capacity; substitutes telehealth for outpatient care when necessary; waitlists individuals when capacity is unavailable; and determines length of stay.
\item \textbf{Individual treatment state update} (Fig.~\ref{fig:A3}): daily processing of treatment assignments, waitlist status and resubmission, treatment completion, stepdown between levels of care, and post-treatment mutual help.
\end{itemize}

AUDTxSim includes four types of parameters: (1) epidemiologic parameters that link demographics with alcohol use and engagement with medical care; (2) natural history parameters that govern transitions between alcohol use and AUD states as a function of personal characteristics and history; (3) intervention parameters that specify how SBIRT, MAUD, mutual help, and facility-based treatment affect state transitions; and (4) treatment facility parameters that define the types of treatment each facility offers, its capacity, and its location. Parameter references in the Appendix~A diagrams (green boxes) index rows of the model parameter file so that each decision node can be traced to its source estimate. We will document the model using the Overview, Design concepts, and Details (ODD) protocol for describing agent-based models \cite{Grimm2020}. A visual ODD summary of the model, following the standardised layout of \cite{Szangolies2024}, is provided in Appendix~B (Fig.~\ref{fig:B1}).

\subsection{Synthetic population and agent characteristics}

AUDTxSim models outcomes for spatially located synthetic populations representing the states of CA and TX \cite{Rineer2025}. Construction of these populations follows the methods documented in \cite{Rineer2025}, integrating Census-based demographic constraints with representative household and person records. Iterative proportional fitting is applied at the Census block-group level to align the synthetic population with local marginal distributions while preserving realistic combinations of individual and household characteristics. Individuals belong to a synthetic population that includes 28,032,218 simulated adults (aged 18 years or older) in CA and 19,325,520 in TX. The population is derived from aggregate Census data and Public Use Microdata Sample (PUMS) records from the 2019 US Census. The baseline dataset includes demographic characteristics for each individual, such as age, race and ethnicity, and sex, and the individual's Census tract of residence. Because each synthetic record is linked to a PUMS record, additional variables---household income, household size, parental and marital status, and employment---are appended. Variables not available through PUMS are added through variable augmentation, including health insurance coverage type, Medicaid eligibility and coverage, educational attainment, and spoken language. Urbanicity is attached to each record through a spatial join. Each individual's location, probabilistically placed according to a population density grid, determines the treatment facilities available to them based on geographic proximity. Finally, each record carries modeled variables derived from these inputs, including current alcohol use or AUD state. Table~\ref{tab:agents} summarizes the attributes of each agent and their use in the model. The model does not account for migration into or out of the study area, nor for natural population growth by birth or aging into the adult population. It does account for mortality: each individual is subject to an annual probability of death that depends on age, sex, race and ethnicity, and current alcohol use and AUD state.

\begin{table*}[t]
\caption{Synthetic population characteristics and their applications in AUDTxSim.}
\label{tab:agents}
\centering\small
\begin{tabularx}{\textwidth}{@{}>{\RaggedRight\arraybackslash}p{2.6cm} L L >{\RaggedRight\arraybackslash}p{2.6cm}@{}}
\toprule
\textbf{Attribute} & \textbf{Description and possible values} & \textbf{Application in AUDTxSim} & \textbf{Source} \\
\midrule
Person ID & Unique numeric identifier & Agent identity & Synthetic population \\
Household ID & Identifier of the household in which the individual lives & --- & Synthetic population \\
State / County / Census tract ID & FIPS and tract codes for place of residence & Predictor of initial alcohol use and AUD state; determines candidate treatment facilities (nearest facilities by service type) & Synthetic population \\
Birth date / current age & Birth date; current age in days updated daily & Predictor of AUD state transitions and death; determines the annual ``birthday'' evaluation day & Synthetic population \\
Sex & Female; Male & Predictor of initial AUD state, transitions, and death; facility preference & Synthetic population \\
Race & American Indian or Alaska Native; Asian; Black or African American; Native Hawaiian or Other Pacific Islander; White & Predictor of AUD state assignment, transitions, and death & Synthetic population \\
Ethnicity & Hispanic or Latinx; Not Hispanic or Latinx & Predictor of AUD state assignment, transitions, and death & Synthetic population \\
Parenthood status & Has children; Does not have children & Predictor of initial AUD state; intervention options and outcomes & Synthetic population (PUMS) \\
Urban/rural residency & Urban; Rural & Intervention options; stratification of outcomes & Spatial join \\
Health insurance status & Uninsured; Medicaid; Other insurance & Healthcare engagement; facility preference and eligibility & Variable augmentation \\
Spoken language, education, employment & Categorical & Facility preference (language services); predictors of engagement & Variable augmentation / PUMS \\
Initial alcohol use / AUD state & Seven states (see Table~\ref{tab:states}) & Determines intervention options and subsequent transitions & NSDUH; NESARC-III \\
Current AUD state (dynamic) & Seven living states plus Deceased & Determines intervention options and subsequent transitions & NESARC-I \& II; initial state \\
Flags (dynamic) & In treatment; Waitlist; Remission; Mutual Help; Brief Intervention; Standalone MAUD & Govern eligibility for annual events and override natural transitions & Model determination \\
\bottomrule
\end{tabularx}
\par\smallskip
\footnotesize NSDUH, National Survey on Drug Use and Health. NESARC, National Epidemiologic Survey of Alcohol and Related Conditions.
\end{table*}

\subsection{Alcohol use and AUD states}

Heavy alcohol use is defined using guidelines for daily (more than 3 drinks per day for women and 4 drinks per day for men) and weekly (more than 7 drinks per week for women and 14 drinks per week for men) consumption \cite{NIAAAweb}. AUD is defined according to DSM-5 \cite{APA2013}, ranging from mild (symptoms in 2--3 of 11 domains) to moderate (4--5 domains) and severe (6 or more domains). AUDTxSim includes eight disease states, seven of which describe a living individual's current status defined jointly by alcohol consumption level and AUD severity (Table~\ref{tab:states}).

\begin{table}[t]
\caption{Alcohol use and AUD states in AUDTxSim.}
\label{tab:states}
\centering\small
\begin{tabularx}{\columnwidth}{@{}>{\RaggedRight\arraybackslash}p{3.1cm} L@{}}
\toprule
\textbf{State} & \textbf{Definition} \\
\midrule
1. No or light (low-risk) alcohol use & No heavy use, no current AUD, no history of AUD \\
2. Heavy alcohol use & Exceeds NIAAA daily or weekly limits; no current AUD or history of AUD \\
3. Mild AUD & Current DSM-5 AUD, 2--3 symptom domains \\
4. Moderate AUD & Current DSM-5 AUD, 4--5 symptom domains \\
5. Severe AUD & Current DSM-5 AUD, 6 or more symptom domains \\
6. Remission with no or light alcohol use & History of AUD, no current AUD, low-risk or no drinking \\
7. Remission with heavy alcohol use & History of AUD, no current AUD, drinking above NIAAA limits \\
8. Deceased & Absorbing state \\
\bottomrule
\end{tabularx}
\end{table}

\subsubsection{Assigning initial states}
At the outset of the simulation every individual is assigned to one of the seven living states in two stages. First, each individual's current alcohol use or AUD state, without regard to AUD history, is drawn from a multinomial logistic regression model predicting current state from demographic characteristics. Second, history of AUD is drawn from a multinomial logistic regression model predicting AUD history given demographic characteristics and current state; individuals without current AUD but with a positive history are placed in one of the two remission states. Both models are estimated from NSDUH and NESARC-III data.

\subsubsection{Transitions among states}
\begin{figure*}[t]
\centering
\includegraphics[width=0.85\textwidth]{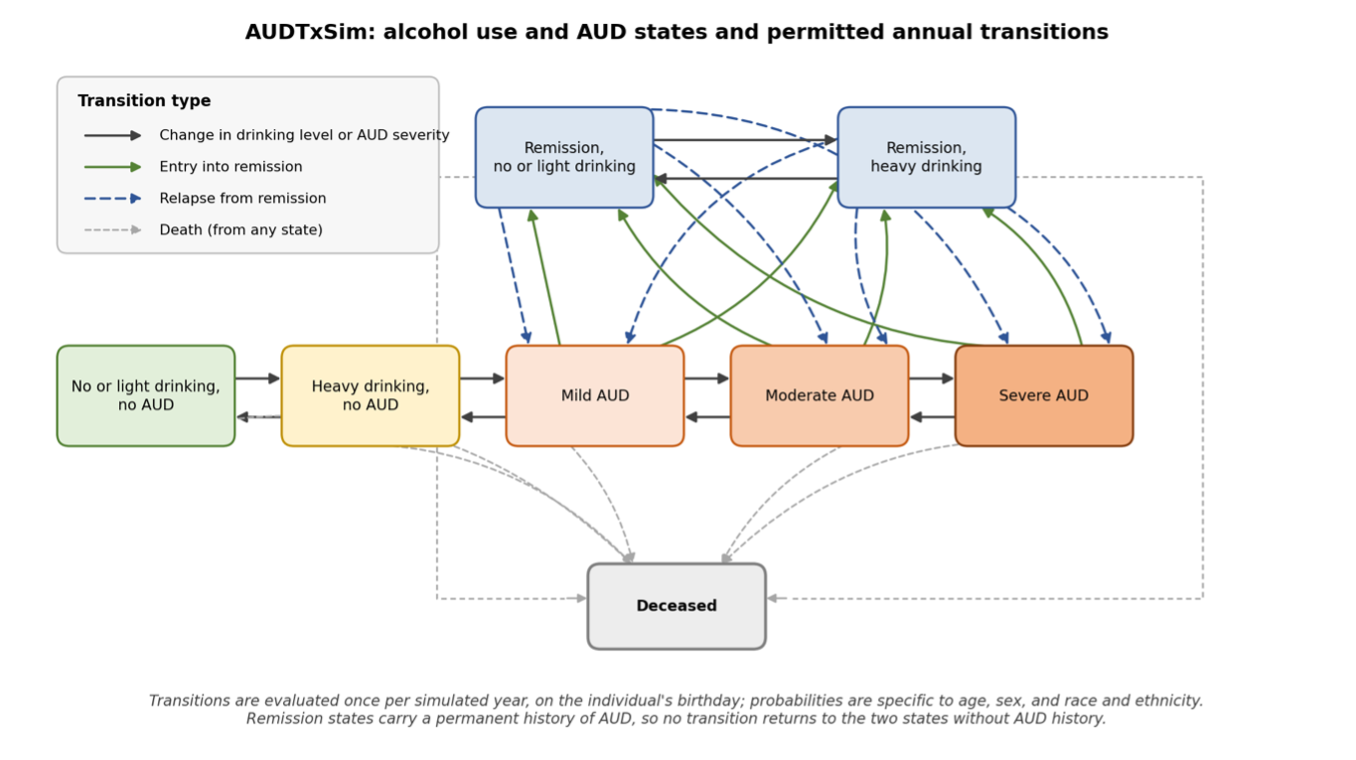}
\caption{Alcohol use and AUD state transitions in AUDTxSim.}
\label{fig:states}
\end{figure*}

Over time an individual's state may change. Annual transition probabilities between alcohol use and AUD states were estimated from the National Epidemiologic Survey on Alcohol and Related Conditions (NESARC) waves 1 and 2, which is the only national survey with longitudinal data on AUD clinical course and treatment utilization \cite{Grant2015,Hasin2015}. Transition probabilities were calculated using multi-state Markov models (R package \texttt{msm} \cite{Jackson2011}), and they are specific to age, sex, and race and ethnicity. Fig.~\ref{fig:states} illustrates the permitted transitions, which together represent the natural history of AUD in the population.

\subsection{Component 1: Annual individual events and brief intervention}

On each simulated day, every individual is checked to determine whether it is their randomly assigned birthday. If not, no annual processing occurs for that individual on that day. If it is their birthday, the individual passes through the sequence of decisions shown in Fig.~\ref{fig:A1}, which we describe here in the order they are evaluated by the model.

\subsubsection{Mortality}
The first annual event is a mortality draw. The probability of dying in the coming year is a function of age, sex, race and ethnicity, and current alcohol use or AUD state, so that individuals in heavy use and AUD states carry excess mortality relative to those in the low-risk state. An individual who dies transitions to the Deceased state; no further events are evaluated, and if the individual was in treatment their facility capacity is released by the treatment manager (Component 2).

\subsubsection{Treatment and waitlist exclusion}
Individuals who are currently receiving facility-based treatment or who are on a waitlist do not undergo natural disease-state transitions, healthcare visits, or screening during their annual evaluation; their disease trajectory during the current year is governed instead by treatment (Component 3). This prevents an individual from, for example, spontaneously relapsing during a residential stay whose outcome has already been determined.

\subsubsection{Mutual help retention}
If the individual is currently engaged in mutual help (e.g., Alcoholics Anonymous or another 12-step program), the model draws whether they remain in mutual help for another year. If they do not, the Mutual Help flag is cleared. In either case, the individual proceeds to the disease-state transition logic.

\subsubsection{Disease-state transition logic}
The model first checks whether the individual had a successful brief intervention or a successful course of standalone MAUD during the preceding year (indicated by the Brief Intervention and Standalone MAUD flags set in the previous annual cycle). If so, the intervention effect is realized: the individual transitions deterministically to the no or light alcohol use state (or, for individuals with an AUD history, to remission with no or light use), and the natural transition is bypassed for that year.

Otherwise, the individual undergoes a natural disease-state transition. If the individual is not in short-term remission, the new state is drawn from the default age-, sex-, and race and ethnicity--specific annual transition probability table (Table~\ref{tab:states}, Fig.~\ref{fig:states}). If the individual is in short-term remission, the transition is governed by a remission transition logic that (i)~tracks time since remission onset; (ii)~records the remission context---natural, unassisted recovery versus recovery through treatment or mutual help; and (iii)~computes a time-varying weight $w(t)$ that interpolates between two relapse risks: the AUD-to-AUD persistence risk associated with the individual's last AUD state, and the long-term remission-to-AUD risk associated with their current remission state (no or low-risk drinking versus heavy drinking). This weighted probability determines whether the individual relapses to AUD or remains in remission. The weight parameters are currently placeholders pending calibration and literature review \cite{Dawson2008,Witkiewitz2020}.

After the new state is drawn, the model classifies the transition. If it is a remission transition (from an AUD state to a non-AUD state), the Remission flag is set and the remission clock starts. If it is a relapse transition (from a remission state back to an AUD state), the individual is removed from mutual help (if applicable) and the Remission flag is cleared. All individuals then proceed to the healthcare and SBIRT logic.

\subsubsection{Healthcare visit, screening, and detection}
Each individual has an annual probability of at least one healthcare visit, which depends on demographic characteristics and insurance status. Individuals without a healthcare visit skip screening and brief intervention; if they have AUD, they may nevertheless seek treatment on their own (see Section~\ref{sec:seeking}), but otherwise no further action is taken.

Individuals with a healthcare visit have a probability of being screened for alcohol use that depends on their characteristics and on the SBIRT implementation scenario being simulated. If screening does not occur, the individual is routed to the treatment-seeking logic if they have AUD. If screening occurs, the model determines whether heavy use or active AUD is detected. In the current model version, detection is deterministic: screening detects the individual's true state if they are in the heavy alcohol use state or in any AUD state (this step is not separately parameterized). Individuals who screen negative take no further SBIRT action; if they have AUD they are routed to treatment seeking regardless of whether they are screened. Screening alone has no effect on subsequent state transitions; its role is to identify candidates for brief intervention and treatment referral.

\subsubsection{Brief intervention}
Among individuals who screen positive for heavy alcohol use, a brief intervention is always considered appropriate by design (a decision node is present so that appropriateness criteria could be introduced in a later model version). The model then draws whether a brief intervention actually occurs, with a probability that reflects the documented gap between screening and brief counseling in primary care. If the brief intervention occurs, its success is drawn with probability of 1 in 7, based on meta-analytic estimates of the number needed to treat for brief alcohol interventions \cite{Kaner2018}. A successful brief intervention has an effect only for individuals in the heavy alcohol use state (i.e., without current AUD). For these individuals, the Brief Intervention flag is set, and at their next birthday they transition to the no or light alcohol use state. Based on the extant literature, brief intervention is not modeled as being successful for individuals already in an AUD state; for them, the relevant intervention pathway involves referral to treatment.

\subsubsection{Referral to treatment}
In parallel with the brief intervention branch, individuals who screen positive for an AUD state are eligible for referral to treatment (referral is considered appropriate only for individuals with AUD; this check is deterministic). Whether a referral actually occurs is a probability that depends on individual characteristics and the SBIRT scenario being modeled. Regardless of whether a referral occurred, all individuals with AUD proceed to the treatment-seeking decision, so that a referral raises but does not solely determine the probability of seeking treatment.

\subsubsection{Treatment seeking, standalone MAUD, and mutual help}
\label{sec:seeking}
Every individual with AUD reaching this point---whether via a healthcare visit with or without referral, or with no healthcare visit at all---draws whether they seek treatment in the coming year. The probability is higher for individuals who received a referral.

Individuals who do not seek formal treatment may receive standalone medication for AUD (MAUD) from a prescriber outside the specialty treatment system. If they receive standalone MAUD, the model draws whether it is successful (probability of 1 in 11, based on meta-analyses of the number needed to treat for MAUD \cite{Jonas2014,McPheeters2023b,McPheeters2023a}). A successful course sets the Standalone MAUD flag, which produces a transition to the no or light use (or remission) state at the next birthday via the disease-state transition logic described above. Individuals who neither seek treatment nor receive standalone MAUD take no further action for the year.

Individuals who seek specialty alcohol treatment first draw whether they also receive standalone MAUD in conjunction with treatment seeking; if they do, a medication add-on is guaranteed in any subsequent facility-based treatment. The model then determines whether the treatment sought is mutual help only. If so, the individual enters standalone mutual help: the Mutual Help flag and the Remission flag are set, no facility request is generated, and retention in mutual help is re-evaluated at each subsequent birthday.

\subsubsection{Preparation of a facility-based specialty treatment request}
For individuals who seek specialty alcohol treatment, the model prepares a treatment profile before any facility is contacted:

\begin{itemize}
\item \textbf{Treatment type.} The required initial level of care is drawn from among hospitalization, inpatient (residential) treatment, and outpatient treatment, with probabilities that depend on AUD severity and individual characteristics.
\item \textbf{Treatment add-ons.} The model draws whether the treatment episode includes medication for AUD and whether it includes concurrent mutual help. The medication add-on is retained only if the eventually assigned facility offers MAUD (pharmacotherapy or medication-assisted treatment); otherwise it is removed by the treatment manager (Component 2).
\item \textbf{Treatment outcome: completion.} Whether the individual will complete the treatment episode is drawn at the time of the initial treatment request, with probabilities informed by the state-specific (California or Texas, approximated by the surrounding southern states due to missing data for Texas) Treatment Episode Data Set (TEDS) discharge records stratified by treatment type and individual characteristics \cite{TEDS}. Determining completion before treatment begins allows the length of stay to be conditioned on both treatment type and treatment outcome.
\item \textbf{Treatment outcome: step down.} Conditional on completion, the model draws whether the individual will step down to a lower level of care (hospitalization to residential; residential to outpatient) or complete the treatment cycle at the current level.
\end{itemize}

With these parameters defined, a formal Treatment Request is generated and submitted to the treatment manager, which processes it the same day.

\subsection{Component 2: Treatment manager handling of new requests and completions}

The treatment manager is a single system-level agent that links patient demand with facility supply (Fig.~\ref{fig:A2}). On each day it performs two operations: it processes treatment completions and then processes new and pending treatment requests.

\subsubsection{Processing completions}
For each Treatment Completion Request received that day (generated by individuals reaching their last day of treatment, Component 3) and for each in-treatment death, the manager releases the occupied slot and updates the facility's occupancy for the relevant treatment type.

\subsubsection{Processing new and pending requests}
For each Treatment Request, the manager applies a facility preference algorithm to select candidate facilities that offer the requested treatment type and match the individual's preferences (e.g., insurance type accepted, sex-specific programming, language services). Candidates are drawn from the individual's pre-computed set of nearest facilities of each service type (see Section~\ref{sec:facilities}), and a choice probability is computed for each candidate conditional on both the individual's characteristics and facility attributes. Facilities are ranked into first-choice and non-first-choice sets.

The manager then applies two feasibility checks in sequence:

\begin{itemize}
\item \textbf{Is at least one matching facility found?} If no facility offers the requested type within the individual's search region and the request is for outpatient care, the treatment type is changed to telehealth and the search is repeated against facilities offering telehealth. If the request is not for outpatient care and no facility is found, no further action is taken on the request.
\item \textbf{Does at least one of the identified (matching) facilities have available capacity?} If all matching facilities are full and the request is for outpatient care, the treatment type is again changed to telehealth and the search is repeated. If the request is for hospitalization or residential care and no capacity exists, the individual is placed on a waitlist. While the individual remains on the waitlist, the request remains active and the manager attempts to fulfil it again on subsequent days (see Component 3 for waitlist duration and resubmission rules).
\end{itemize}

When at least one facility with capacity is available, the manager randomly selects one of the first-choice facilities; if none have availability, it then performs a weighted random selection among non-first-choice facilities using the choice probabilities. Next it verifies whether the selected facility offers MAUD; if it does not, any medication add-on in the individual's treatment profile is removed unless the individual received standalone MAUD outside of the formal treatment system. Length of stay is then drawn from state-level length-of-stay distributions specific to the treatment type (derived from N-SUMHSS) and to the pre-determined completion outcome derived from the TEDS-discharge data. Telehealth episodes currently draw from the outpatient length-of-stay distribution. Finally, a Treatment Assignment specifying facility, treatment type, add-ons, and length of stay is issued to the individual, and the facility's occupied capacity for that modality is incremented. Facility occupancy and the set of facilities at capacity are logged daily.

\subsection{Component 3: Individual treatment state update}

At the beginning of each individual's simulated day, before any annual events, the individual's treatment status is updated according to the logic in Fig.~\ref{fig:A3}.

\subsubsection{Admission}
If the individual holds a Treatment Assignment issued by the treatment manager on the previous day, they enter facility-based treatment. The In-Treatment flag and the Remission flag are set to true, and the treatment-days-remaining counter is initialized to the assigned length of stay. While in treatment the individual is exempt from natural disease-state transitions and SBIRT events occurring on their birthday (Component 1).

\subsubsection{Waitlisting and resubmission}
If the treatment manager waitlisted the individual during their last treatment request, the Waitlist flag is set. On each subsequent day the model checks whether the individual has been on the waitlist for the maximum waitlist duration (currently 49 days). If so, the individual is removed from the waitlist and discontinues their treatment search for the year; they return to the natural-history and treatment-seeking logic at their next birthday. Otherwise, if the current day falls on the waitlist resubmission interval (every 7 days), a new Treatment Request with the same profile is generated and sent to the manager; on other days no action is taken.

\subsubsection{Discharge, completion, and stepdown}
For individuals in treatment, the model checks whether today is their last day of treatment. If not, no further action is taken. If so, a Treatment Completion Request is generated and sent to the manager to release capacity, and the model processes discharge according to the modality completed:

\begin{itemize}
\item \textbf{Hospitalization.} If the pre-determined outcome includes a step down, a new Treatment Request for inpatient (residential) care is generated. Otherwise the completion status (completed or not completed) is recorded.
\item \textbf{Inpatient (residential).} If a step down was determined, a new Treatment Request for outpatient care is generated. Otherwise completion status is recorded.
\item \textbf{Outpatient and telehealth.} Completion status is recorded; no further step down is available.
\end{itemize}

Stepdown requests are processed as new treatment requests. Their add-ons, completion outcome, and any further stepdown are drawn anew rather than carried over from the completed episode, and they are subject to the same facility matching, telehealth substitution, and waitlist rules. Completion versus non-completion is determined at request time (Component 1); it is recorded here so that it can be tracked and used as a predictor in future decisions---for example, differential remission and relapse probabilities for completers and non-completers.

\subsubsection{Post-treatment mutual help}
After any treatment episode concludes without a subsequent step down, the model draws whether the individual continues into mutual help. If so, the Mutual Help flag is set, and annual retention is re-evaluated at each subsequent birthday (Component 1). Regardless of the mutual help decision, the In-Treatment flag is set to false upon completion of treatment. The Remission flag set at admission remains active, so that the individual enters the remission transition logic---with treatment recorded as the remission context---at their next birthday.

\subsection{Treatment facilities}
\label{sec:facilities}

A comprehensive dataset of AUD treatment facilities in Texas and California was generated from the Substance Abuse and Mental Health Services Administration's (SAMHSA) treatment locator database (findtreatment.gov). Facilities offering alcohol use treatment were selected and classified according to their services; facilities not offering alcohol services were excluded. Facilities located up to 300 miles outside state borders were retained to account for individuals who travel out of state. Facility attributes (Table~\ref{tab:facilities}) include the treatment modalities offered (hospitalization, inpatient/residential, outpatient, and telehealth), availability of MAUD, insurance types accepted, sex-specific programming, and language services.

To approximate service regions, a routing database was created from population-weighted Census tract centroids to all facilities within 300 miles using the OpenStreetMap road network to compute driving distance and travel time. Each individual was assigned the ten closest facilities of each service type based on their tract of residence; these constitute the candidate set of treatment programs searched by the treatment manager.

Each facility has a state-specific (California or Texas) total capacity for each treatment modality it offers, an occupied capacity (initialized from estimated current census), and a resulting available capacity used by the treatment manager. Occupied and available capacities are dynamic and change daily as patients are admitted and discharged; total capacity is static within a run but can be scaled by modality-specific capacity multipliers to represent capacity-expansion simulation scenarios. Capacities for different modalities within a facility are independent and not interchangeable (e.g., slots designated for hospitalized patients cannot be used for residential patients, or vice versa).

\begin{table*}[t]
\caption{Facility attributes.}
\label{tab:facilities}
\centering\small
\begin{tabularx}{\textwidth}{@{}>{\RaggedRight\arraybackslash}p{2.8cm} L L@{}}
\toprule
\textbf{Attribute} & \textbf{Description and possible values} & \textbf{Application in AUDTxSim} \\
\midrule
Sex & Sex of individual a facility serves \{male, female\} & Used in preference calculation for an individual \\
Insurance type & Insurance type(s) a facility accepts: \{private, Medicare, Medicaid\} & Used in preference calculation for an individual \\
Language & Services offered in \{English, Spanish, Asian languages, Other\} & Used in preference calculation for an individual \\
Type & \{Outpatient, Inpatient, Hospitalization\} & Model assigned treatment type for an individual \\
Total capacity & Max values: \{Outpatient: 79, Inpatient: 34, Hospitalization: 20\} & Determines the total number of patients a facility can accept \\
Occupied capacity & 0 to total capacity by treatment type & The number of patients at a facility at any given time step \\
Waitlist & 0--99 & If facility is full, individuals are placed on waitlist \\
Location & Latitude, longitude (location on the road network) & Reachability for an individual is determined by the road network where actual travel distance is less than 300 miles (rural) or 50 miles (urban) \\
\bottomrule
\end{tabularx}
\end{table*}

\subsection{Data sources}

Multiple data sources are used to construct agents, to parameterize the decision nodes described above, and to calibrate the model (Table~\ref{tab:data}). Where possible, parameters are estimated separately by state and by demographic stratum. Where state-specific data are unavailable, national estimates or peer-reviewed literature are used. All final parameter values will be reported with the model results to ensure transparency and replicability.

\begin{table*}[t]
\caption{Data sources used for model inputs, parameterization, and calibration.}
\label{tab:data}
\centering\small
\begin{tabularx}{\textwidth}{@{}L L@{}}
\toprule
\textbf{Model element} & \textbf{Data source} \\
\midrule
Agent demographics and location (age, sex, race/ethnicity, household, tract, income, employment, education, insurance, language) & RTI SynthPop synthetic population, 2019 Census/PUMS base with variable augmentation \cite{Rineer2025} \\
Initial alcohol use and AUD state; AUD history & NSDUH (2019--2023); NESARC-III \\
Annual natural-history transitions between states & NESARC waves 1 and 2, multi-state Markov models (\texttt{msm}) \\
Remission and relapse dynamics & NESARC; peer-reviewed literature \cite{Dawson2008,Witkiewitz2020} \\
Mortality by state and demographics & CDC WONDER / state vital statistics; alcohol-attributable relative risks from the literature [to confirm] \\
Healthcare visit, screening, brief intervention, referral probabilities & NSDUH (past-year healthcare use and alcohol screening/counseling items) \\
Brief intervention success (1 in 7) & Meta-analyses and systematic reviews of brief alcohol interventions \\
Treatment seeking, treatment type, add-ons, completion, stepdown & NSDUH; Treatment Episode Data Set (TEDS) admissions and discharges \cite{TEDS} \\
Standalone MAUD receipt and success & NSDUH; peer-reviewed literature [to confirm] \\
Mutual help entry and annual retention & NSDUH; peer-reviewed literature [to confirm] \\
Length of stay by modality and completion status & TEDS discharges, state-level \\
Facility locations, services, MAUD, capacity, insurance, language & SAMHSA findtreatment.gov; N-SSATS/N-SUMHSS [to confirm capacity source] \\
Travel distance and time to facilities & OpenStreetMap road network routing from tract centroids \\
Calibration targets: healthcare visits, heavy drinking prevalence, mortality & NSDUH state and substate estimates; BRFSS; CDC WONDER \\
\bottomrule
\end{tabularx}
\end{table*}

\subsection{Technical implementation}

The model is implemented in Python using Repast4Py \cite{Collier2022}, the Python member of the Repast suite of agent-based modeling toolkits. Repast4Py builds on the message passing interface (MPI) standard (via \texttt{mpi4py}) to support agent-based models distributed across the processes (``ranks'') of a high-performance computing cluster. We use three components of the toolkit: its agent abstraction (\texttt{repast4py.core.Agent}), its shared agent population container (\texttt{repast4py.context.SharedContext}), and its discrete-event scheduler (\texttt{repast4py.schedule}). The simulation advances in daily time steps, and a parameters file processed through Repast4Py's parameters module supplies all model inputs, supporting reproducible, configuration-driven runs.

\subsubsection{Individual agents}
Each member of the synthetic population is represented by an individual agent that subclasses \texttt{repast4py.core.Agent} and is identified by an (id, type, rank) tuple. Individuals are instantiated from a synthetic population microdata file with demographic attributes, health attributes (current disease state, comorbid mental health and substance use indicators, and AUD history), and treatment-related states (in treatment, treatment days remaining, assigned facility, waitlist status, and cumulative treatment counts). At each daily step, the scheduler invokes the individuals' update behavior (Component 3), during which each agent reports its current state for logging; processes any treatment assignment returned by the manager and enters or continues treatment or remains on a waitlist, accordingly; completes treatment when its remaining duration elapses; and re-issues treatment requests if still waiting. On each agent's birthday step the agent ages, is subjected to mortality, attempts a natural-history transition, and passes through the SBIRT and treatment-seeking logic (Component 1). Each agent retains a longitudinal record of its annual disease states and of the steps at which state transitions occurred, which is serialized to columnar (Parquet) output at the end of the run.

\subsubsection{Treatment manager}
Allocation of individuals to treatment is handled by a single \texttt{Treatment\_manager} agent of a distinct agent type. Because treatment capacity is a shared, contended resource that must be allocated consistently across the whole population, exactly one instance is created, on rank 0. The manager holds the full set of facilities for the modeled geographic state (e.g., California or Texas), which is loaded from a facilities file, optionally scaled by capacity multipliers, with each detailing its service offerings, per-modality capacity (outpatient, inpatient, hospitalization, telehealth), and current occupancy. On each step, the manager frees capacity for treatment completions and in-treatment deaths and then processes new and pending requests for treatment as described in Component 2.

\subsubsection{Parallel execution}
The model is launched with \texttt{mpirun} over a configurable number of ranks, each running the same model code over \texttt{MPI.COMM\_WORLD}. The synthetic population is partitioned across ranks in a preprocessing step so that each rank loads only its own subset of individuals into a local \texttt{SharedContext}. Two repeating daily events are registered: the per-individual update on every rank, and the manager update, which is registered with \texttt{PriorityType.LAST} so that it runs after all individuals have been updated. Because capacity is allocated centrally but individuals are distributed across processing nodes, the two events are coordinated by explicit MPI message passing rather than by spatial projections or agent migration. After updating its local individuals, each rank sends its treatment requests, completions, and in-treatment deaths to rank 0 using tagged point-to-point messages terminated by a sentinel. Rank 0 gathers these messages, records which rank owns each individual, invokes the manager to free and allocate capacity globally, and returns each assignment to the owning rank. Individuals act on assignments at the start of the following step. This design confines the contended resource to a single coordinating agent while allowing the population update---the dominant computational cost---to scale across ranks, and keeps each individual on a single rank for the entire run so that no ghost-agent synchronization is required.

\subsection{Parameterization, calibration, and validation}

Each decision node in Figs.~\ref{fig:A1}--\ref{fig:A3} is linked to a row of the model parameter file, and parameters are specified as conditional probabilities that may depend on individual attributes and current state. Before scenario analysis, the model will be calibrated so that simulated healthcare visit rates, heavy drinking prevalence, and mortality reproduce observed values at the state, rural--urban, sex, race and ethnicity, age-group, and county levels, to ensure that observed differences between states, between rural and urban residents, and between sexes are reproduced by the model. The calibration approach is summarized in the Discussion and detailed in a separate calibration plan; all pre-specified criteria and any deviations will be documented in a calibration report. Validation will proceed in three steps: confirming that mortality dynamics of the agent population reproduce state vital statistics; comparing simulated trends in prevalence, treatment receipt, and mortality under the status quo scenario with empirical estimates; and sensitivity analyses varying key parameters to test the robustness of model conclusions to alternative assumptions. Throughout, the study team will qualitatively assess whether model behavior is consistent with the clinical and health-services literature.

\section{Discussion}

\subsection{Summary}

This protocol describes the AUD Treatment Simulation model (AUDTxSim), a stochastic, spatially explicit, individual-based simulation of alcohol use and AUD, and the associated care cascade (including SBIRT and specialty treatment) for the adult populations of California and Texas. The model represents every adult in the two states as an agent geo-located to a Census tract and linked to nearby treatment facilities; simulates annual transitions among eight alcohol use and AUD states, including remission and relapse dynamics that depend on remission duration and context; models the cascade from healthcare visit to screening, brief intervention, referral, and treatment seeking; represents standalone MAUD and mutual help as alternatives and adjuncts to facility-based care; allocates treatment requests to facilities through a centralized manager that respects preferences, geography, and finite modality-specific capacity, substituting telehealth for outpatient care and waitlisting individuals when capacity is unavailable; and follows individuals through admission, stepdown between levels of care, discharge, and post-treatment mutual help. The daily time step and the birthday-triggered annual events allow the model to represent both slow disease dynamics and fast treatment-system dynamics in a single framework, and the Repast4Py implementation allows populations of tens of millions of individuals to be simulated on a high-performance computing cluster.

To our knowledge, no previous simulation model has coupled the natural history of AUD, the SBIRT cascade, and a capacity-constrained, geographically explicit treatment system at the scale of full state populations. Because access to primary care, screening, and treatment differs across racial and ethnic, insurance, language, and rural--urban groups, and because treatment capacity is finite and unevenly distributed, expanding SBIRT reach may not translate into proportional gains in treatment receipt and may widen, rather than narrow, disparities. Once calibrated, AUDTxSim will allow us to quantify these dynamics: how much of the potential benefit of universal screening is lost to limited healthcare engagement; how much is lost to downstream capacity constraints, waitlists, and telehealth substitution; which subgroups benefit most and least; and whether targeted improvements in access or capacity are more effective than untargeted expansion.

\subsection{Calibration plan}
\label{sec:calib}

Calibration will target three outcomes for which reliable observed data exist: healthcare visit rates, heavy drinking prevalence, and mortality. Targets are defined at the state level (CA, TX), by rurality (rural, urban), by state-by-rurality cell, by sex, by race and ethnicity (White non-Hispanic, Black non-Hispanic, Hispanic, Other), by age group, and at the county level, with combinations of strata to be explored later. The first step is an assessment of the current state: using existing baseline runs for both states, observed and simulated values for each target and stratum will be compared after a two-year burn-in. Relative differences will be tabulated to identify the most pressing calibration needs and to define accuracy that is ``good enough.'' The guiding criterion is that calibration should reproduce the observed differences between states, between rural and urban settings, and between sexes, because these contrasts are central to the study aims. For example, if the observed difference in an AUD-related outcome between CA and TX is 15\%, the calibrated model should reproduce each state's value within roughly 7\% and the between-state difference within 10--20\% of its observed magnitude.

Because each full-population run of a ten-year simulation is computationally expensive, simulator evaluations are treated as a scarce resource, and the plan extracts multiple uses from every run. Rather than the 500--1,000 runs per scenario and top-5--10\% posterior retention envisioned for the most comprehensive calibration design, the current calibration plan uses 40--60 initial space-filling runs followed by sequential refinement; replaces posterior retention with a history-matching (admissible-region) approach supplemented by weighted likelihood or surrogate-assisted Bayesian updating; uses one fixed development sample with common random numbers followed by selected independent replicates; reports annual outputs through year 10 with selected diagnostics; and propagates uncertainty into scenarios through a small, stratified ensemble of representative parameter sets rather than a full posterior. A parsimonious target vector spanning the causal pathways (healthcare engagement, drinking, mortality) is preferred over many correlated individual statistics, and a limited set of demographic strata is chosen for scientific importance and data reliability.

To make calibration tractable, comparisons will first be conducted at the county level. Two counties---one urban and one rural---will be selected in each state, giving a four-way comparison whose runs are inexpensive enough to be conducted on laptops or modest cloud instances, with the number of runs determined by observed effect sizes. Once these four counties are calibrated, the same comparison will be repeated for four different counties to confirm that the parameterization generalizes, providing internal validation. Finally, a small number of full-state runs (perhaps one per state) will confirm that the calibrated parameters reproduce the statewide CA and TX targets.

The fit criterion for each target $j$ is a standardized discrepancy
\begin{equation}
z_{j}(\theta)=\frac{y_{j}(\theta)-m_{j}}{s_{j}},
\label{eq:discrepancy}
\end{equation}
where $y_{j}(\theta)$ is the simulated estimate, $m_{j}$ the empirical target, and $s_{j}$ combines target uncertainty, a pre-specified model-discrepancy allowance, and estimated Monte Carlo error. Squared relative error is acceptable initially, with cosine similarity considered for the vector-valued target. Stochastic variation will be managed by using common random numbers for design runs, estimating Monte Carlo variance from strategically placed replicates (baseline, one poor-fit point, and several candidate best-fit points), favoring unique design points early and replication later, and storing compact run-level summaries rather than complete agent histories.

The calibration steps are as follows: (1)~pre-calibration---checking whether existing CA and TX runs already reach the three targets at state and county levels, extreme-value tests at parameter bounds to confirm directional behavior, baseline stochastic runs to quantify Monte Carlo noise, and locking the transition matrix, target definitions, target-data version, and parameter setup before the first formal design; (2)~parameter selection---sensitivity analysis at the local level to identify which parameters (initially expected to be the natural-history transition parameters) are most useful for calibration, followed by a maximin Latin hypercube design of about 50 runs over transformed inputs respecting correlations and logical constraints, with preliminary regressions or tree models fit after 20--30 runs for quality control only; and (3)~final confirmation---fitting multiple candidate surrogates, running about 30 new runs where surrogates predict low discrepancy, high uncertainty near the admissibility boundary, or poor coverage, selecting a small diverse set of finalist parameter vectors, and using a final 20 runs as independent-seed replications reporting means, Monte Carlo standard errors, target residuals, and pass/fail status. A reference parameter set will be chosen only after this confirmation, and an ensemble retained for uncertainty propagation. Outputs will include a parameter summary (estimated, fixed, or scenario-only), observed-versus-simulated plots with uncertainty intervals, residual heat maps by state, subgroup, and year, the plausible parameter ensemble with a non-identifiability statement, and a calibration report documenting pre-specified criteria and all deviations.

\subsection{Limitations}

AUDTxSim has several limitations. First, it simulates two of the 50 U.S. states, and findings may not generalize to states with different insurance landscapes or treatment systems, although the mechanisms modeled---access-dependent screening, capacity-constrained treatment, and heterogeneous natural history---are not specific to CA and TX. Second, several parameters, notably remission and relapse weights, the effect of treatment completion on subsequent trajectories, and standalone MAUD and mutual help effectiveness, rest on limited data and are currently placeholders or literature-based estimates. Effects of variation in these will be examined in sensitivity analyses. Third, the model does not represent migration, births, or aging into the adult population, and treatment outcomes are pre-determined at request time rather than emerging from within-treatment dynamics. Fourth, detection upon screening is treated as perfect, and brief intervention effectiveness is applied only to individuals without AUD. Finally, the model necessarily simplifies a complex system. The calibration plan, ODD documentation, and sensitivity analyses are intended to make these simplifications transparent and their consequences quantifiable.

\section{Conclusion}

Fewer than one in twelve adults with AUD receive treatment, and access to the primary care encounters where heavy drinking is most often identified is unequal. AUDTxSim provides a detailed, data-driven, population-scale platform for exploring how expanding SBIRT reach and treatment capacity would change AUD prevalence, treatment receipt, and mortality in California and Texas, and for whom. By describing the model's structure, decision logic, data sources, implementation, and calibration plan in advance, this protocol establishes the basis on which the model's eventual findings should be evaluated.

\section*{Acknowledgments}

The authors gratefully acknowledge the work of Caroline Kroma to the development of the custom synthetic populations for this project. Michael Long also contributed to early model development and programming. We also thank the expert advisory panel members who have contributed to the refinement of the simulation model throughout the development process.

\printbibliography

\appendix
\renewcommand{\thesection}{\Alph{section}}
\renewcommand{\thefigure}{A\arabic{figure}}
\setcounter{figure}{0}
\section*{Appendix A: Full Model Logic Diagrams}

The following diagrams (version dated 18 September 2026) specify the decision logic implemented in AUDTxSim. Green boxes reference rows of the model parameter file (Statistics and Modeling / Modeling / ModelParameters / Current Model Parameters.xlsx); purple boxes are implementation notes; bold-bordered boxes are terminal nodes for the page.

\begin{sidewaysfigure*}
\centering
\includegraphics[width=\textheight,height=\textwidth,keepaspectratio]{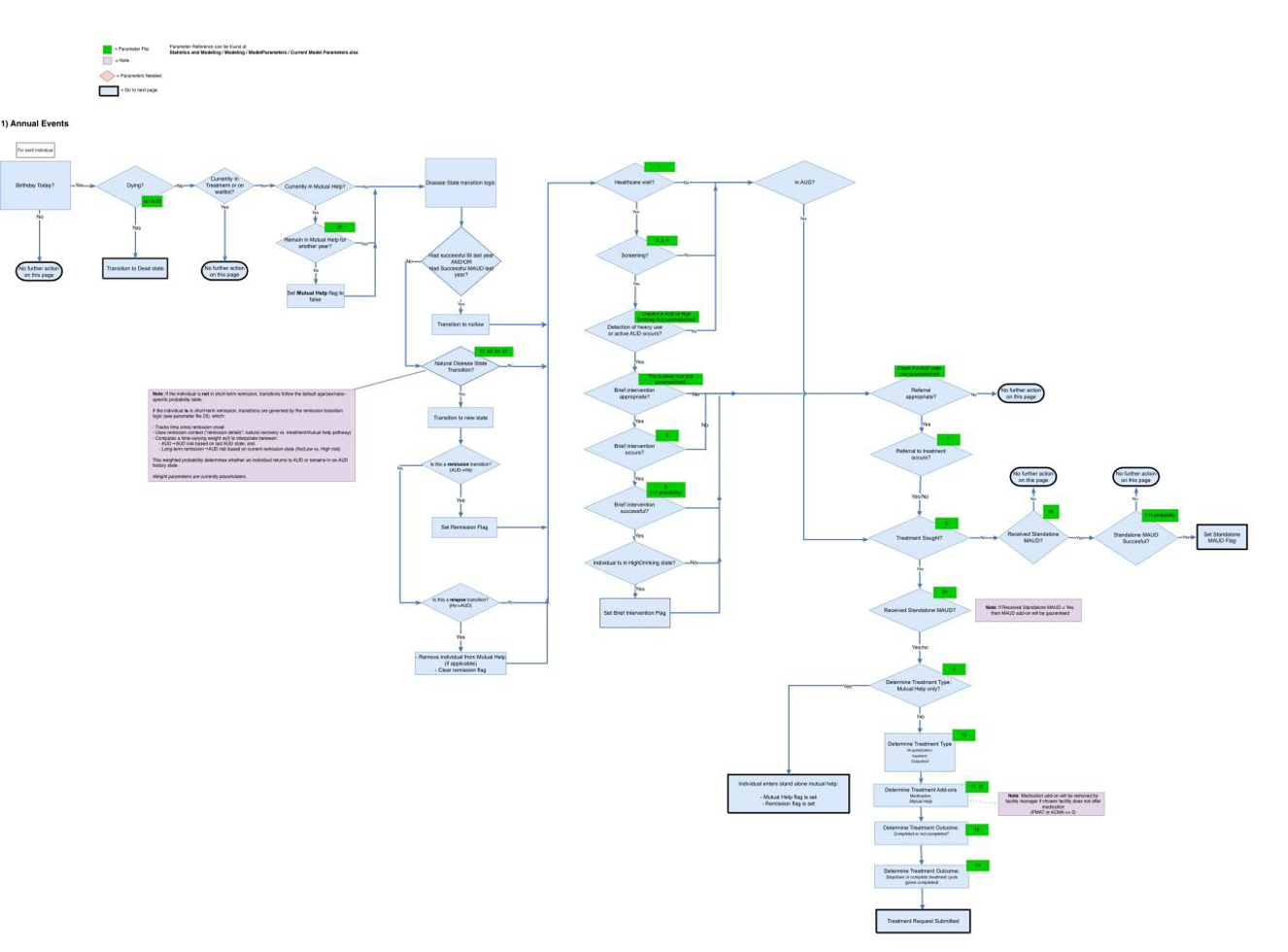}
\caption{Component 1 --- Annual individual events and brief intervention.}
\label{fig:A1}
\end{sidewaysfigure*}

\begin{figure*}[p]
\centering
\includegraphics[width=\textwidth,height=0.92\textheight,keepaspectratio]{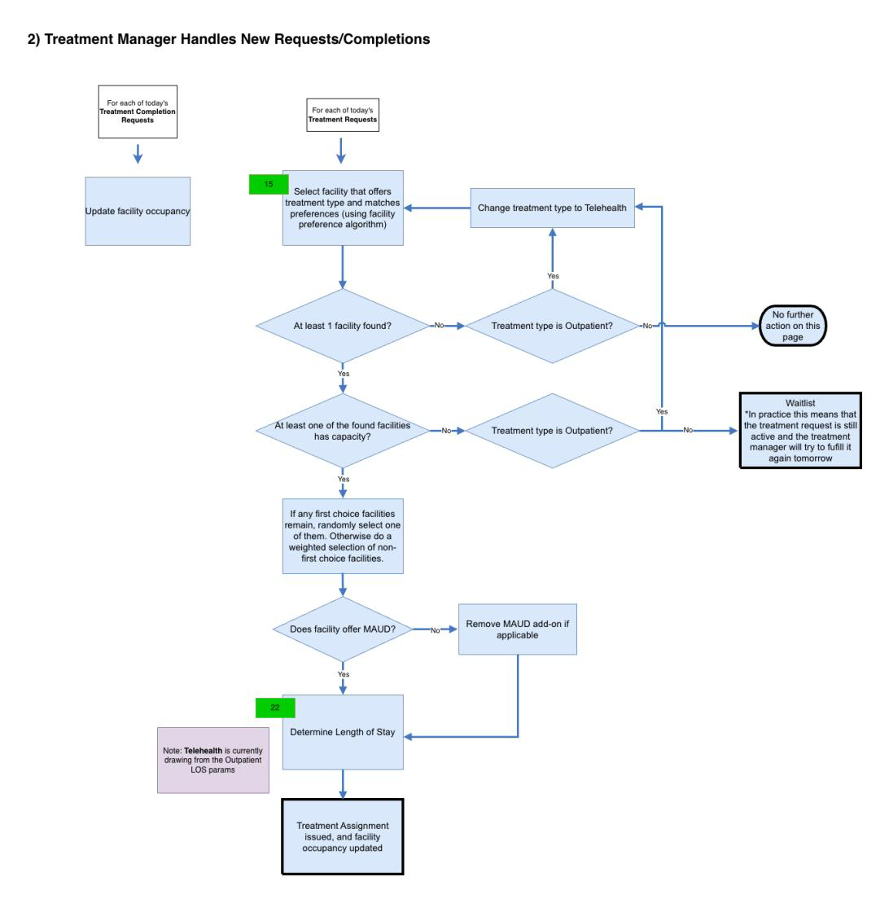}
\caption{Component 2 --- Treatment manager handling of new requests and completions.}
\label{fig:A2}
\end{figure*}

\begin{sidewaysfigure*}
\centering
\includegraphics[width=\textheight,height=\textwidth,keepaspectratio]{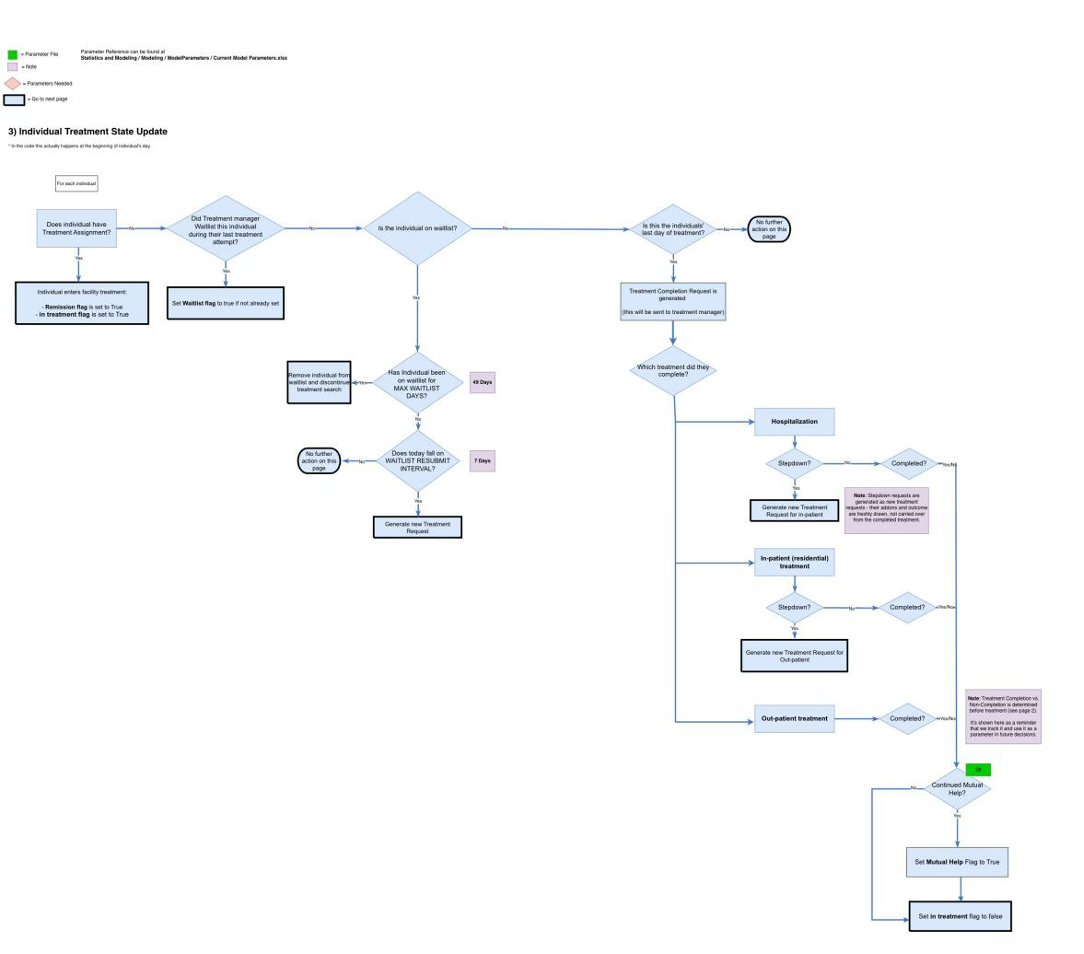}
\caption{Component 3 --- Individual treatment state update.}
\label{fig:A3}
\end{sidewaysfigure*}

\section*{Appendix B: Visual ODD}
\renewcommand{\thefigure}{B\arabic{figure}}
\setcounter{figure}{0}

\begin{sidewaysfigure*}
\centering
\includegraphics[width=\textheight,height=\textwidth,keepaspectratio]{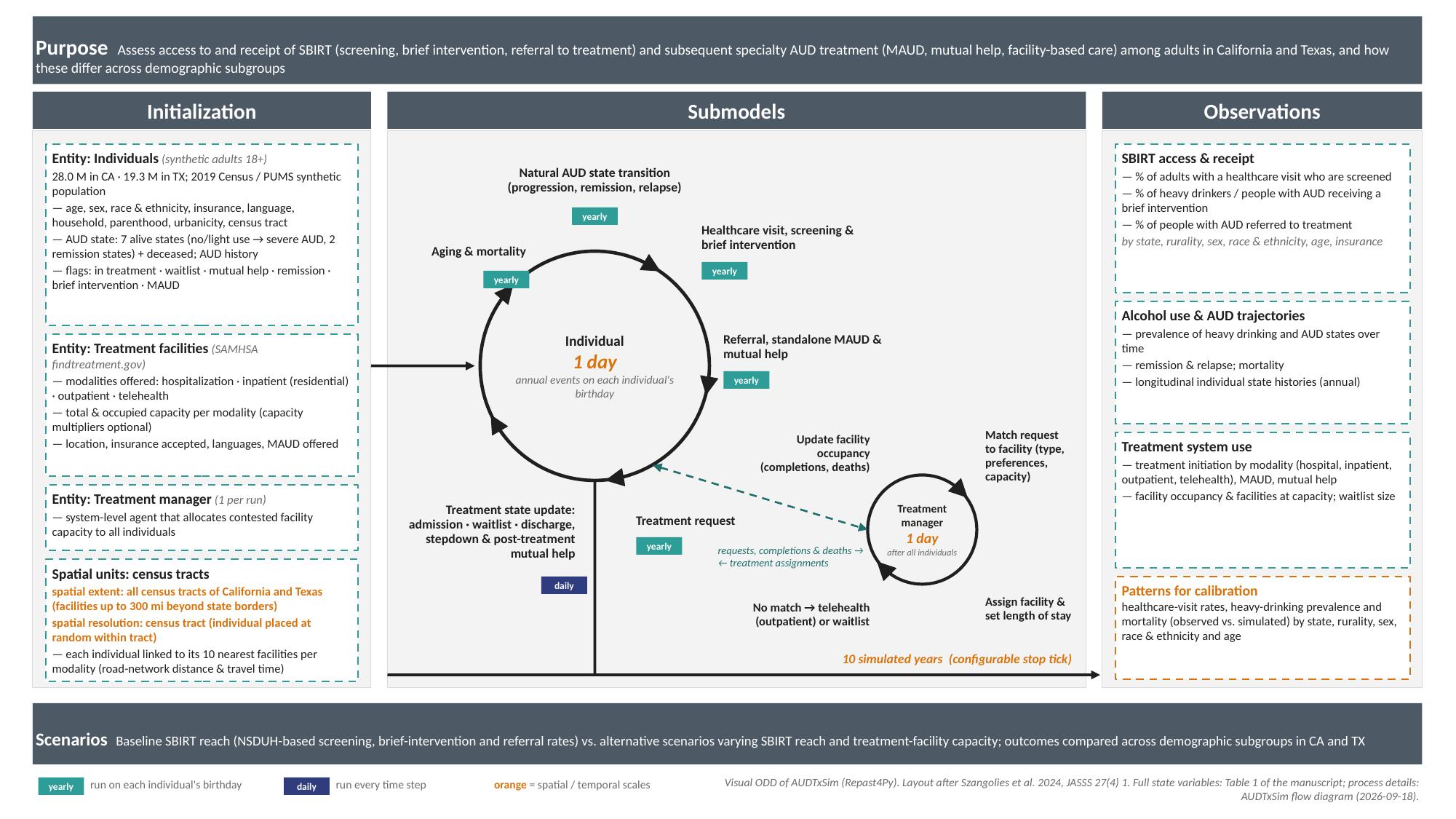}
\caption{Visual ODD of AUDTxSim.}
\label{fig:B1}
\end{sidewaysfigure*}

Fig.~\ref{fig:B1} summarizes AUDTxSim as a visual ODD \cite{Szangolies2024}, which arranges the model's purpose, initialisation (entities and spatial units), submodels and their scheduling, observations, and scenarios in a standardised layout.

\end{document}